\documentclass[%
reprint,
superscriptaddress,
amsmath,amssymb,
aps,
prb,
]{revtex4-2}

\usepackage{booktabs}
\usepackage{multirow}
\usepackage{color}
\usepackage{graphicx}
\usepackage{dcolumn}
\usepackage{bm}
\usepackage{hyperref}
\usepackage[mathlines]{lineno}
\usepackage{tabularx}

\begin{document}
	
	
	\title{Light-engineered Multichannel Quantum Anomalous Hall Effect in High-order Topological Plumbene}
	
	\author{Zhe Li}%
	\email{lizhe21@iphy.ac.cn}
	\affiliation{%
		Beijing National Laboratory for Condensed Matter Physics, and Institute of Physics, Chinese Academy of Sciences, Beijing 100190, China
	}%

	\author{Fangyang Zhan}
	\affiliation{%
		Institute for Structure and Function, Department of Physics, and Chongqing Key Laboratory for Strongly Coupled Physics,
		Chongqing University, Chongqing 400044, People’s Republic of China
	}%

	\author{Haijun Cao}
	\affiliation{%
		Beijing National Laboratory for Condensed Matter Physics, and Institute of Physics, Chinese Academy of Sciences, Beijing 100190, China
	}%
	
	\affiliation{%
		University of Chinese Academy of Sciences, Beijing 100049, China
	}

	\author{Jingjing Cao}
	\affiliation{%
		College of Physics and Electronic Information, Shanxi Normal University, Taiyuan, 030006, China
	}%

	\author{Huisheng Zhang}
	\email{hszhang@sxnu.edu.cn}
	\affiliation{%
		College of Physics and Electronic Information, Shanxi Normal University, Taiyuan, 030006, China
	}%
	
	\affiliation{%
		College of Chemistry and Materials Science $\&$ Key Laboratory of Magnetic Molecules and Magnetic Information Materials of Ministry of Education $\&$ Research Institute of Materials Science, Shanxi Normal University, Taiyuan, 030006, China
	}

	\author{Sheng Meng}
	\email{smeng@iphy.ac.cn}
	\affiliation{%
		Beijing National Laboratory for Condensed Matter Physics, and Institute of Physics, Chinese Academy of Sciences, Beijing 100190, China
	}%
	
	\affiliation{%
		University of Chinese Academy of Sciences, Beijing 100049, China
	}
	
	\affiliation{%
		Songshan Lake Materials Laboratory, Dongguan, Guangdong 523808, China
	}
	
	\date{\today}
	
	\begin{abstract}
		Floquet engineering severs as a forceful technique for uncovering high Chern numbers of quantum anomalous Hall (QAH) states with feasible tunability in high-order topologically insulating plumbene, which is readily accessible for experimental investigations. Under the irradiation of right-handed circularly polarized light, we predict a three-stage topological phase transition in plumbene, whether it is in a free-standing form or grown on h-BN. Initially, a metallic state evolves into a $K$($K'$)-valley-based QAH state with a Chern number of --8, which then decreases to --6 after the valley gap closes. Finally, a band inversion occurs at the $\Gamma$ point, resulting in a multichannel QAH state with $C$ = --3. The trigonal warping model accounts for both $K$($K'$)-valley-based and $\Gamma$-point-based QAH states. Additionally, growing plumbene on a non-van-der-Waals substrate eliminates the $K$($K'$)-valley-based topology, leaving only the $\Gamma$-point-based QAH state with $C$ = +3. Our findings propose the tunability of various high Chern numbers derived from high-order topological insulators, aiming to advance the next-generation dissipationless electronic devices.
	\end{abstract}
	
	\maketitle

	\section{Introduction}


	The theoretical predictions, computational simulations, and experimental fabrications of two-dimensional (2D) topological insulators have sparked a new wave of research in the past two decades \cite{geim2013van,novoselov20162d,huang2017layer,burch2018magnetism,zak1990universal,zhong2017van,buscema2014fast,haldane1988model,kane2005quantum,molle2017buckled,balendhran2015elemental,ezawa2015monolayer,tang2025unveiling,liu20222d,grazianetti2024future,pan2014valley,pan2015valley,li2024multimechanism,xue2024valley,hong2025designing,vila2021valley,bernevig2006quantum,hasan2010colloquium,qi2011topological,fu2007topological,xu2013large,reis2017bismuthene,liu2011stable}. Among the numerous candidate materials, graphene, along with low-buckled group-IV Xenes, plays a pivotal foundational role in 2D topology. The initial prediction of the quantum anomalous Hall (QAH) effect, proposed by F. D. M. Haldane, is grounded in graphene \cite{haldane1988model}. Similarly, in graphene, Kane and Mele proposed a 2D quantum spin Hall (QSH) system characterized by a tiny gap located at two valleys \cite{kane2005quantum}. The enhancement of the spin-orbit coupling (SOC) effect increases this gap to the meV scale, while maintaining the QSH features in other members of the same group: silicene, germanene, and stanene \cite{molle2017buckled,balendhran2015elemental,ezawa2015monolayer,tang2025unveiling,liu20222d,grazianetti2024future}. In these free-standing films, spatial inversion symmetry ($P$) is preserved along with valley degeneracy, in the absence of external influences. However, breaking $P$ introduces Rashba terms that lift valley degeneracy, resulting in various valley-polarized topologies, including the quantum valley Hall state and valley-polarized QAH state, among others \cite{pan2014valley,pan2015valley,li2024multimechanism,xue2024valley,hong2025designing,vila2021valley}. The valley-polarized topology can also be achieved by growing Xenes on suitable substrates \cite{li2024multimechanism,xue2024valley,hong2025designing,vila2021valley}. These findings expand the family of 2D topological materials, showcasing novel properties.

	Floquet engineering offers a remarkably powerful tool for the continuous manipulation of valley-based topologies. The light-induced Chern insulating state has been extensively investigated through both theoretical and experimental analyses \cite{liu2023floquet,zhan2024perspective,oka2009photovoltaic,bao2022light,liu2018photoinduced,hubener2017creating,zhan2023floquet,zhu2023floquet,qin2023light,wan2024photoinduced,liu2022high,li2025light,zhang2025quantum,wang2013observation,mahmood2016selective,choi2025observation,merboldt2025observation,zhou2023pseudospin,qiu2018ultrafast,bielinski2025floquet,mciver2020light,usaj2014irradiated,liu2023floquet-b,li2024floquet,tian2025quantized,bukov2015universal,blanes2009magnus}. Silicene, germanene, and stanene, which exhibit visible nontrivial valley gaps, can undergo a topological phase transition (TPT) from QSH to QAH states, with the latter supporting operation beyond room temperature \cite{li2025light}. Additionally, several 2D valley-based magnets demonstrate similar phenomena but with only one spin component. For example, monolayer ScI$_2$ and VSNH undergo a two-stage TPT process: starting from a second-order topological insulator, first transitioning to a QAH state, then evolving into a normal insulating state as the light intensity increases \cite{li2024floquet,tian2025quantized}. Linearly polarized light (LPL) preserves valley degeneracy, while circularly polarized light (CPL) establishes polarity between the two valleys. Monolayer VSi$_2$N$_4$ exhibits both a high Chern number and valley polarity under CPL irradiation, which is contributed by the trigonal warping model \cite{zhan2023floquet}. Overall, Floquet engineering effectively excavates novel topological properties hidden within these systems.
	
	The last member of group-IV Xene, plumbene, loses its 2D topological manifestation but instead exhibits high-order topology, despite possessing a significant global gap \cite{zhang2021selective}. This property stems from the internal friction between $K$($K'$)-valley-based and $\Gamma$-point-based topologies. Leveraging these insights, we systematically extract the Chern insulating characteristics of plumbene through Floquet engineering in this study. Our findings reveal, for the first time, a multiple-momentum-based multichannel QAH (mQAH) state (the chiral edge states develop at multiple momentum points) \cite{li2024multimechanism} within a single monolayer system, based on band inversions between $\pm$1 orders of replica bands. Specifically, both the $K$($K'$)-valley-based QAH state, as described by the 	Haldane model \cite{haldane1988model,li2024multimechanism,xu2022controllable}, and the $\Gamma$-point-based QAH state, characterized by the Qi-Wu-Zhang model \cite{li2024multimechanism,xu2022controllable,qi2006topological}, coexist while exhibiting opposite chirality, both further complicated by the trigonal warping model \cite{fu2009hexagonal,liu2013plane,ren2016quantum,zhong2017plane}. Under right-handed CPL (R-CPL) irradiation, a complex sequence of TPTs occurs, beginning from a metallic state and advancing through QAH states with high Chern numbers of --8, --6, and --3 in succession. Notably, growing plumbene on a non van der Waals (non-vdW) substrate effectively filters out the $K$($K'$)-valley-based QAH state, leaving only the $\Gamma$-point-based one ($C$ = +3). Given the existing experimental fabrication of plumbene on various substrates \cite{yuhara2019graphene,mihalyuk2025plumbene,bihlmayer2020plumbene}, our proposal of the mQAH state in a single material presents a richer selection of Chern numbers. This allows for the selective imposition of TPTs on either the $K$($K'$)-valley or $\Gamma$-point-based topology, underscoring the potential for diversified signal selection in future electronic devices.
	

	\section{Computational Methods}
	
	The ground states of plumbene with its heterostructures grown on various substrates were acquired employing first-principles computational methods implemented in the Vienna ab-initio Simulation Package (VASP) \cite{kresse1996efficient}. An 11$\times$11$\times$1 $k$-mesh grid was adopted for the relaxation and self-consistent calculations across all these building-blocks. We adopted the criterion of structural optimization, ensuring that the Hellmann-Feynman force on each atom was less than 0.001 eV/Å. For the convergence criterion of the electron energy, a threshold of 1.0 $\times$ $10^{-7}$ eV was established for relaxations, self-consistent calculations, and band structure computations. For performing the electronic-correlation, the PBE functional was utilized to calculate all the ground states \cite{perdew1996generalized}. During the structural relaxation step, the SOC effect was shut down; nonetheless, it was included in the self-consistent and band structure computation phases. Moreover, a vacuum layer of at least 15 Å along the out-of-plane direction was incorporated to simulate the 2D behavior \cite{tkatchenko2009accurate} among these systems. The vdW corrections \cite{grimme2010consistent} were not included for free-standing plumbene and plumbene/non-vdW-substrate, but imposed to systems of plumbene/vdW-substrate. The finite displacement method was utilized to obtain the phonon spectra. After ground-state computation, the PHONOPY \cite{togo2015first} and VASPKIT \cite{wang2021vaspkit} codes were employed for the post-processing of the obtained data.
	
	To investigate the excited states under periodic optical fields, we initially employed the Wannier90 package to construct the tight-binding Hamiltonians (TBHs) for all relevant materials, utilizing maximally localized Wannier functions \cite{mostofi2014updated, marzari1997maximally, souza2001maximally}. Following the acquisition of the ground-state TBHs, we implemented a code developed by our group to apply the Peierls substitution \cite{oka2019floquet}: 
	
	\begin{equation}
		H_{ij}(t) = H_{ij} \exp\left[ -\mathrm{i} \int_{R_j}^{R_i} \textbf{\textit{A}}(t) \cdot d\mathbf{r} \right]
		\label{eq1}
	\end{equation}
	
	In eq. \ref{eq1}, $R_i$ and $R_j$ stand for the two sites of Wannier charge center.
	
	Once we had acquired the Floquet-engineered TBH, we utilized the Green's function method through the WannierTools package \cite{wu2018wanniertools} to explore the topological characteristics of edge states, Berry curvature distributions, plane-resolved gap distributions, and Chern numbers. The $z2pack$ code in Chern-number mode was employed to further validate the Chern number in each topological phase by calculating the distributions of Wannier charge centers \cite{gresch2017z2pack}.
	
	Finally, we list the code implementation of eq. \ref{eq1}. Specifically, the displacement vector can be defined as:
	
	\begin{equation}
		\begin{split}
			l_{ij} = L_k+w_i-w_j
		\end{split}
		\label{eq2}
	\end{equation}
	
	In eq. \ref{eq2}, $w_{i}$ and $w_{j}$ are related to the two sites of Wannier charge center. $L_k$ is the Cartesian coordinate of the $k$-th lattice point.  
	
	The amplitude of the optical vector potential can be expressed as:
	
	\begin{equation}
		\begin{split}
			A_{ij} & = \Biggl[  \left( A_x l_{ij,x} \sin \phi_1 + A_y l_{ij,y} \sin \phi_2 + A_z l_{ij,z} \sin \phi_3 \right)^2 \\
			& + \left( A_x l_{ij,x} \cos \phi_1 + A_y l_{ij,y} \cos \phi_2 + A_z l_{ij,z} \cos \phi_3 \right)^2 \Biggr]^{1/2}
		\end{split}
	\end{equation}
	
	$A_x$, $A_y$ and $A_z$ are light intensities along three axes, while $\phi_1$, $\phi_2$ and $\phi_3$ are phase parameters of the light.

	The phase angle $\phi_{ij}$ forms as:
	
	\begin{equation}
		\phi_{ij}  = 
		\tan^{-1}\left( \frac{A_x l_{ij,x} \sin \phi_1 + A_y l_{ij,y} \sin \phi_2 + A_z l_{ij,z} \sin \phi_3}{A_x l_{ij,x} \cos \phi_1 + A_y l_{ij,y} \cos \phi_2 + A_z l_{ij,z} \cos \phi_3}  \right)
	\end{equation}
	
	After the excitation of the light, the expanded Hamiltonian can be expressed as:
	
	\begin{equation}
		H_{ij}^{(q)} = H_{ij}^{(k)} \textit{e}^{\mathrm {i} q \phi_{ij}} J_q \left( \frac{e}{\hbar} A_{ij} \right)
	\end{equation}

	$H_{ij}^{(k)}$  represents the original Hamiltonian obtained from the Wannier functions, while $J_q$
	denotes the $q$-th order Bessel function that characterizes the tunneling effect facilitated by photons. The variable $q$ indicates the number of photons that are either emitted or absorbed. In this study, $q$ is varied through the values --2, --1, 0, +1, and +2. The $\pm2$ order replica bands are not present in the final band structures; they are included solely to introduce perturbations to the zero-order bands and the $\pm1$ order replica bands.

	\section{Results}
	
	\subsection{Ground States}
	
	Free-standing plumbene manifests a significantly stronger SOC compared to stanene, resulting in a notable $K$($K'$)-valley gap of 400 meV. Figure S2 in Supplementary Materials \cite{supplementary} confirms its trivial insulating behavior in the first-order regime ($Z_2$ = 0). The corner states clearly verify its high-order topology, depicted in Fig. S2(c). Notably, the topological edge state connects the valence band at both the $K$($K'$)-valley and the $\Gamma$-point, underscoring its internal friction characteristics. It is indeed regrettable that such a large gap Xene, which harbors topologically nontrivial potentials, succumbs to internal dissipation. These hidden first-order topologies become dominant under the Floquet engineering regime. Regarding its significant size of gap, it's hard to observe the TPT of zero-order bands (obtained after Magnus expansion \cite{liu2023floquet,blanes2009magnus,bukov2015universal}) before the light intensity reaches too high. Besides, due to the large gap, ultrahigh light frequency is acquired to exclude the influence of replica bands, which is distant from practical experimental applications. Hence, here we consider $\pm$1-order of replica bands, which undergo post-processing following their hybridization with the zero-order bands and the $\pm$2-order replica bands.


	It is worth mentioning that, at the free-relaxed regime the final topological state transitions into a metallic phase due to the misalignment of the band gap (see Fig. S3 in Supplementary Materials \cite{supplementary}); however, the $K$($K'$)-based chiral edge states still can be unambiguously observed within the gaps. To achieve band gap alignment between the $K$($K'$) valleys and the $\Gamma$ point, a moderate biaxial strain of +3.0\% is applied [see Fig. S4(c) in Supplementary Materials \cite{supplementary}, \textbf{\textit{a}} = 5.08 Å], following methods similar to those used in germanene/MnBi$_2$Te$_4$-family systems \cite{li2024multimechanism,hong2025designing,li2019intrinsic,li2020tunable,tang2023intrinsic}. Additionally, this tensile strain has no effect on the structural stability (Fig. S5). Consequently, the subsequent discussion will primarily focus on this condition, as it facilitates a detailed examination of the TPT and the globally gapped topological characteristics.

	\subsection{Three Stages of Floquet QAH States with High Chern Numbers}
	
\begin{figure*}
		\centering
		\includegraphics[width=1\linewidth]{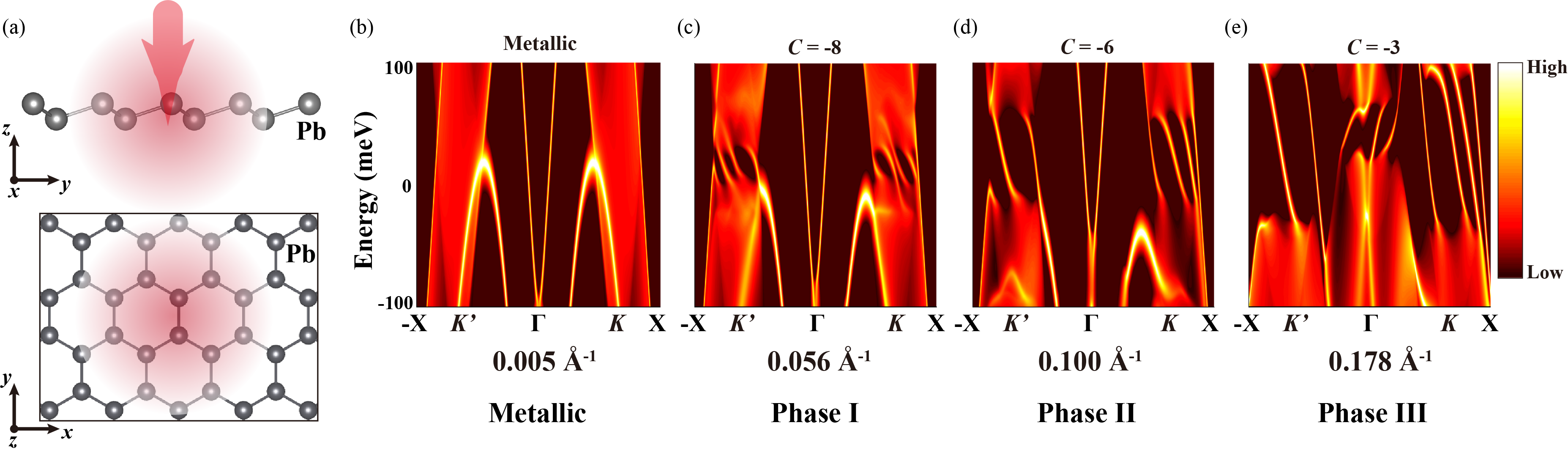}
		\caption{\textbf{The evolution of the LDOS patterns in R-CPL-irradiated free-standing plumbene (at in-plane lattice constant as 5.08 Å and the light frequency of $\hbar\omega$ = 0.375 eV).} (a) shows the side view (upper panel) and top view (lower pannel) of the plumbene under the irradiation of R-CPL, with the gray sphere representing the Pb atom. (b)-(e) provide a detailed presentation of the band inversion process and illustrate the $Y$-edge projected LDOS patterns at light intensities of 0.005 Å$^{-1}$, 0.056 Å$^{-1}$, 0.100 Å$^{-1}$ and 0.178 Å$^{-1}$, respectively. The total Chern number is indicated at the top of each subfigure, while the corresponding topological phase is shown at the bottom (Phases I, II, III stand for the Chern-insulating states with $C = -8$, $C = -6$ and $C = -3$ respectively). In each subfigure, the color gradient progresses from black to red and then to light-yellow, indicating an increase in LDOS values. 
		}
		\label{fig1:LDOS}
	\end{figure*}

	\noindent
	Notably, the TPT process involves more complex intermediate phases. Starting from the local density of states (LDOS) patterns projected onto the $Y$ edge, as shown in Fig. \ref{fig1:LDOS}, we select an optical frequency of $\hbar\omega$ = 0.375 eV and consider four representative light intensities. Figure \ref{fig1:LDOS}(a) illustrates the side view of free-standing plumbene under the irradiation of CPL. At an ultralow light intensity (defined as $A$, $A = \frac{eA_0}{\hbar }$ \cite{liu2023floquet,blanes2009magnus,bukov2015universal,zhu2023floquet,li2023tunable}, here $A$ = 0.005 Å$^{-1}$), almost no band reformation is observed, with significant overlap of bands around the $K$ and $K'$ valleys [Fig. \ref{fig1:LDOS}(b)]. This overlap first disrupts, resulting in a higher Chern number for the valley-based QAH state (Phase I, $C$ = --8) when the light intensity is increased to 0.056 Å$^{-1}$ as demonstrated in Fig. \ref{fig1:LDOS}(c). Subsequently, before the band closure around the $\Gamma$ point, another TPT occurs, reducing the Chern number to --6 (Phase II), as illustrated in Fig. \ref{fig1:LDOS}(d) at $A$ = 0.100 Å$^{-1}$. Above 0.120 Å$^{-1}$, the band around the $\Gamma$ point closes and then reopens, giving rise to a $\Gamma$-point-based QAH state. Figure \ref{fig1:LDOS}(e) clearly reveals this mQAH state (Phase III), characterized by a compensated Chern number ($C$ = --6 + 3 = --3). The Chern numbers for these three topological phases are further confirmed by the anomalous Hall conductance (AHC) spectra and the distributions of Wannier charge centers (Fig. S6). For Phase III, the $K$ ($K'$)-valley-based Chern number ($-6$) and the $\Gamma$-point-based Chern number ($+3$) are individually validated through partial-region integration of the Berry curvatures (Fig. S7). Importantly, the global gap remains intact, with the gap value increasing as the light intensity exceeds 0.150 Å$^{-1}$. This developing behavior suggests complex band interactions arising from different spins and momentum points.
	
\begin{figure*}
		\centering
		\includegraphics[width=0.8\linewidth]{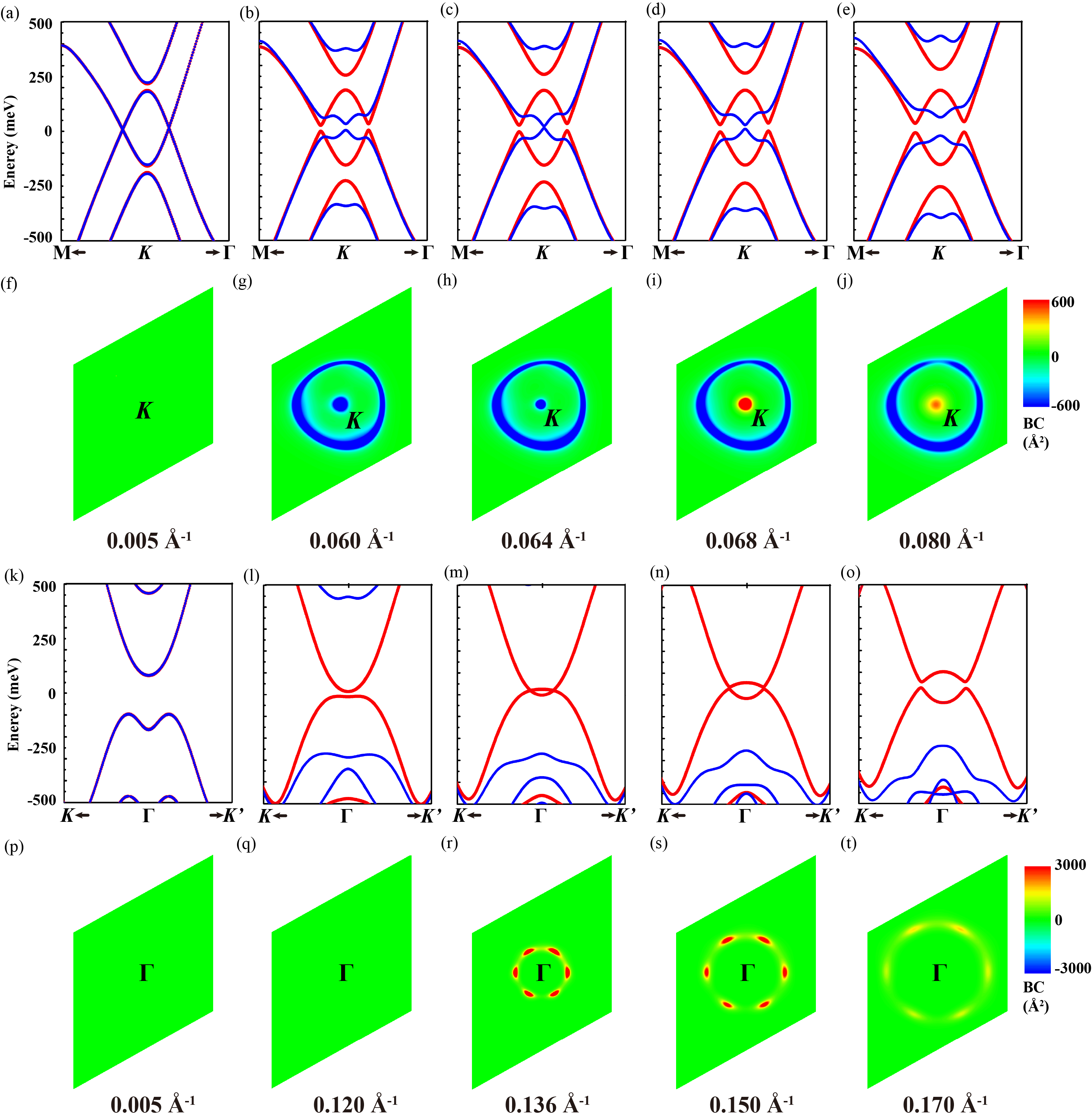}
		\caption{\textbf{A detailed analysis of the TPT process separately at the $K$ valley and the $\Gamma$ point: free-standing plumbene at in-plane lattice constant as 5.08 Å and $\hbar\omega$ = 0.375 eV.} Panels (a)–(e) display the spin-resolved band evolutions around the $K$ valley, while panels (f)–(j) illustrate the zoomed-in Berry curvature distributions in the same region. The selected light intensities are 0.005 Å$^{-1}$, 0.060 Å$^{-1}$, 0.064 Å$^{-1}$, 0.068 Å$^{-1}$ and 0.080 Å$^{-1}$, respectively. In the band structures, the red and blue components represent spin up and down. In the Berry curvature plots, the red and blue regions indicate positive and negative chirality, respectively. Panels (k)–(t) provide a similar analysis around the $\Gamma$ point, with the light intensities chosen as 0.005 Å$^{-1}$, 0.120 Å$^{-1}$, 0.136 Å$^{-1}$, 0.150 Å$^{-1}$ and 0.170 Å$^{-1}$, respectively.}
		\label{fig2:Detail_analysis}
	\end{figure*}

	In further detail, Fig. \ref{fig2:Detail_analysis} elucidates the specifics of the aforementioned three-stage TPT process. Maintaining an optical frequency of $\hbar\omega$ = 0.375 eV, Figs. \ref{fig2:Detail_analysis}(a)–\ref{fig2:Detail_analysis}(e) explore the spin-resolved band developments near the $K$ valley, with spin up and down represented in red and blue, respectively. Since valley degeneracy is preserved due to the presence of $P$, we focus our analysis solely on the $K$ valley, and those of $K'$ valley share the same performance. Specifically, spin up opens a gap around (but not directly at) the $K$ point; its negatively signed Berry curvature distribution exhibits a threefold, $C_{3z}$-rotational symmetry [Figs. \ref{fig2:Detail_analysis}(g)–\ref{fig2:Detail_analysis}(j)], indicative of a trigonal warping nature. In contrast, spin down displays a rather unusual behavior: it undergoes a band closure exactly at the $K$ point at around 0.064 Å$^{-1}$ [Fig. \ref{fig2:Detail_analysis}(c)], accompanied by a sign reversal of the Berry curvature at the $K$ point [Figs. \ref{fig2:Detail_analysis}(h) and \ref{fig2:Detail_analysis}(i)]. This unique band closure signifies the TPT process between Phases I and II, which reduces the Chern number from --8 to --6 (with each valley contributing +1). The plane-resolved band-gap evolutions around the $K$ valley also coincides with the above properties [see Figs. S8(a)-S8(e) of Supplementary Materials \cite{supplementary}]. No additional TPT occurs around the $K$ ($K'$) valley as the light intensity increases to higher values.

	The spin-resolved band evolution around the $\Gamma$ point demonstrates distinct behaviors. Beginning with a gapped structure, the two spin components exhibit opposite trends:  spin up experiences a closure and subsequent reopening of the gap, while spin down shows the monotonically gap-increasing performance. The overlap between the conduction and valence band of spin up occurs between 0.120 Å$^{-1}$ and 0.136 Å$^{-1}$, during which the trigonal warping model offers the intrinsic mechanism, as evidenced by the Berry curvature patterns [Figs. \ref{fig2:Detail_analysis}(r)–\ref{fig2:Detail_analysis}(t)]. In this particular QAH state, a sixfold, positively signed Berry curvature distribution forms, with each zone contributing a Chern number of +0.5. As the light intensity increases, this sixfold Berry curvature distribution expands its $k$-space range, as illustrated in Figs. \ref{fig2:Detail_analysis}(r)–\ref{fig2:Detail_analysis}(t). The plane-resolved gap distributions around $\Gamma$ point also obeys sixfold rotational symmetry [see Figs. S8(f)-S8(j) in Supplementary Materials \cite{supplementary}]. All these developments are consistent with the LDOS patterns discussed in Fig. \ref{fig1:LDOS}. Furthermore, these behaviors are summarized in Figs. \ref{fig3:Contours}(a) (around the $\Gamma$ point) and \ref{fig3:Contours}(b) (around the $K$ valley) by depicting spin-resolved gap evolutions, also at $\hbar\omega$ = 0.375 eV. Conversely to the $\Gamma$ point, where spin up causes band gap closure around 0.125 Å$^{-1}$, the closure for the $K$ ($K'$) valleys occurs at spin down around 0.065 Å$^{-1}$. These two TPT points are ideally staggered, creating a clear series of TPT transitions for each phase.

	\subsection{Phase Diagrams: Selecting Appropriate Laser Parameters}
	
\begin{figure*}
		\centering
		\includegraphics[width=0.87\linewidth]{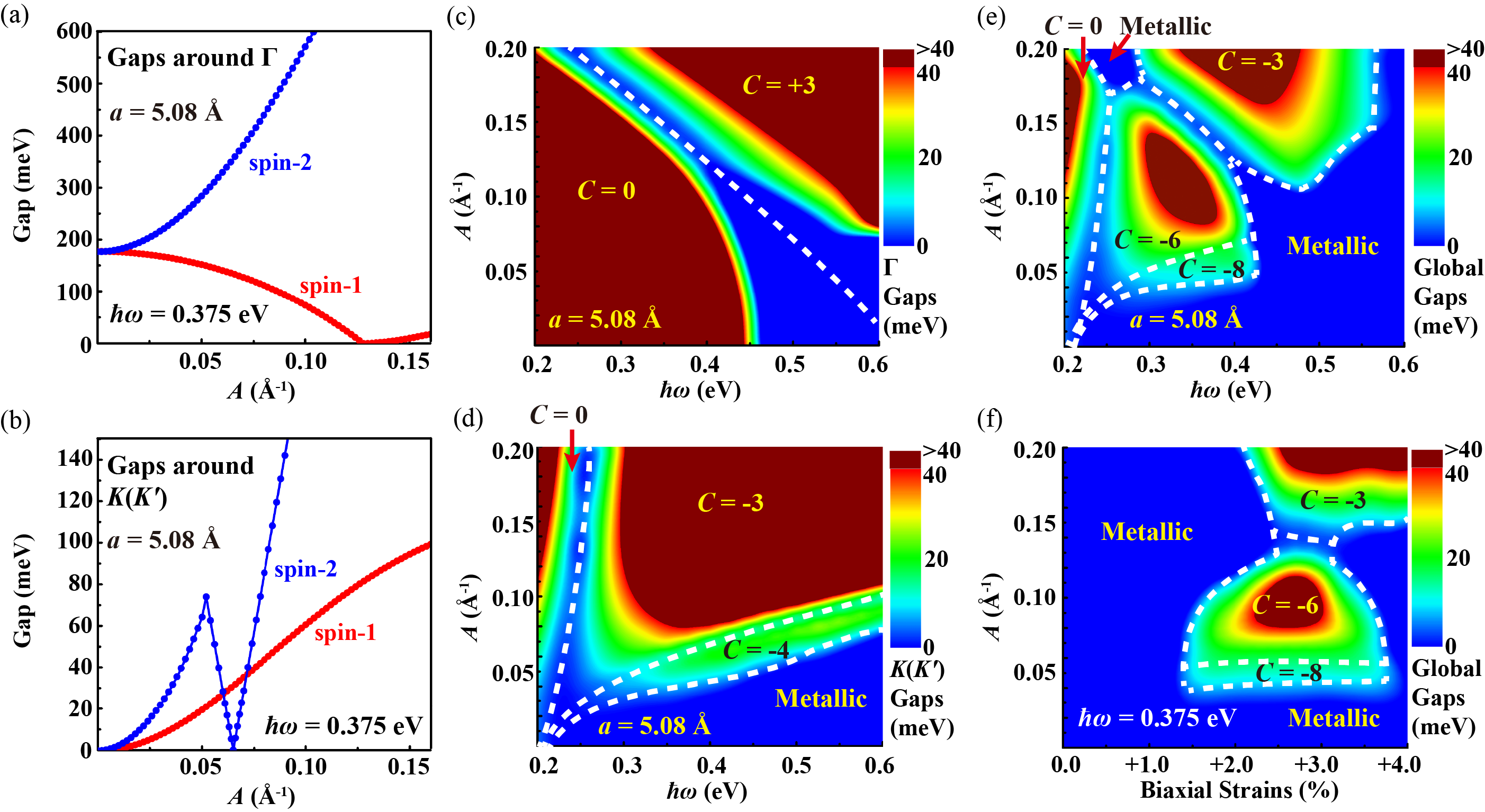}
		\caption{\textbf{Spin splitting and the contour distribution of phase diagrams in R-CPL-induced topologies of free-standing plumbene.} The spin-resolved local gap evolution of free-standing plumbene under R-CPL irradiation at  $\hbar\omega$ = 0.375 eV is shown around (a) the $\Gamma$ point and (b) the $K$($K'$) valley, respectively. The red and blue lines correspond to spin up and down accordingly. The contour distributions of (c) $\Gamma$-point local gaps, (d) $K$($K'$)-valley local gaps, and \textbf{e} global gaps across the entire Brillouin zone (BZ) are depicted as functions of light frequency ($x$-axis) and light intensity ($y$-axis). The color gradient transitions from blue to green and then to red, indicating an enhancement of the global gap. The maroon region represents the global gaps greater than 40 meV, while the metallic region is denoted in blue. The white dashed curves delineate the boundaries between different Chern number phases. Panels (a)-(e) are based on free-standing plumbene with in-plane lattice constant as 5.08 Å. Panel (f) is similar to (e) but illustrates the distributions as a function of biaxial strain ($x$-axis) and light intensity ($y$-axis) at $\hbar\omega$ = 0.375 eV.}
		\label{fig3:Contours}
	\end{figure*}

	The multi-momentum positioned mQAH state in plumbene imposes more stringent requirements on the selection of appropriate laser parameters to achieve a significant global gap feature. Accordingly, Figs. \ref{fig3:Contours}(c) and \ref{fig3:Contours}(d) illustrate the contour distributions of local gaps and Chern numbers around the $\Gamma$ point and $K$ ($K'$) valley, respectively. The case at the $\Gamma$ point presents a straightforward contour shape, with a prominent high Chern-number region ($C$ = +3) located in the upper right area. In contrast, the topological phase distribution around the $K$ ($K'$) valley presents a more complex behavior, predominantly characterized by the QAH state with $C$ = --3. Additionally, a horizontally distributed region with a higher Chern number ($C$ = --4) appears prior to the band closure of spin down corresponding to generally diminished global gaps ($\leq$ 20 meV).
	
	It is gratifying that the final global gap distributions, resulting from the intersection of the two phase-diagrams, reveal two distinct Chern-insulating islands containing ultra-large gapped regions (maroon zones in Fig. \ref{fig3:Contours}(e), $\geq$ 40 meV). Particularly, a quasi-triangular island situated in the left-to-middle portion corresponds to a single $K$ ($K'$) valley-based QAH state, encompassing Phases I and II as mentioned in Fig. \ref{fig1:LDOS}. This island features an ultra-large gapped region that solely supports the phase with $C$ = --6. After traversing a TPT curve characterized by a gapless line, another QAH state island appears at the upper part, related to the mQAH state with $C$ = --3, identified as Phase III in Fig. \ref{fig1:LDOS}. In this region, the maroon zone persists, indicating the potential for the mQAH state to operate beyond room temperature. This marks the first proposed mQAH state triggered by a single pure material with high-temperature operational capability, which represents a significant improvement over the germanene/MnBi$_2$Te$_4$ system \cite{li2024multimechanism}.

	Fig. \ref{fig3:Contours}(f) illustrates the contour distribution of global gaps as a function of biaxial strains (measured in percentage along the $x$-axis), highlighting its essential role in imposing moderate biaxial strains. As expected, no global gap emerges in the free-relaxed condition or at biaxial strains below +1.5\%. The QAH state gradually disintegrates into two islands: the first corresponds to Phases I and II, while the second represents Phase III. As the lattice is stretched, the $K$($K'$)-valley-based QAH state appears first, persisting until approximately +3.8\% biaxial strain, after which it disappears. Notably, the mQAH state emerges later, starting at +2.2\%, and remains stable even beyond +4.0\% biaxial strain. The phase diagram presented here emphasizes the necessity of biaxial strains for effective band gap alignment.

	\begin{figure*}
		\centering
		\includegraphics[width=1\linewidth]{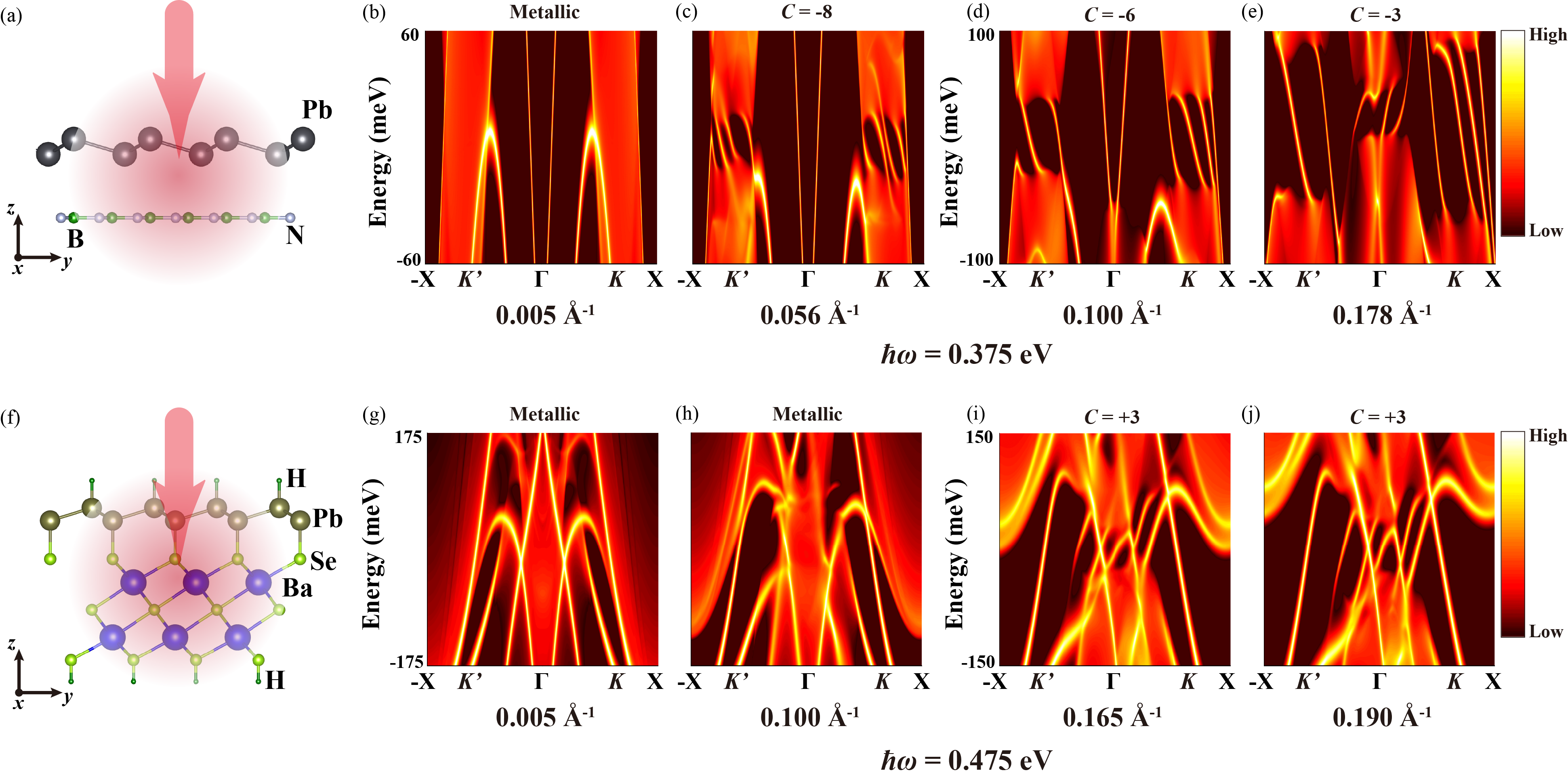}
		\caption{\textbf{R-CPL-induced TPTs of plumbene grown on different substrates.} Panel (a) provides the side view of plumbene/h-BN under the irradiation of R-CPL. The gray, light-gray and green balls stand for Pb, N and B atoms sequentially. Panels (b)–(e) illustrate the $Y$-edge projected LDOS patterns at light intensities of 0.005 Å$^{-1}$, 0.056 Å$^{-1}$, 0.100 Å$^{-1}$ and 0.178 Å$^{-1}$, respectively, with the light frequency as $\hbar\omega$ = 0.375 eV. The total Chern number is indicated at the top of each subfigure, while the corresponding topological phase is noted at the bottom. In each subfigure, the color gradient transitions from black to red and then to light-yellow, reflecting an increase in the LDOS values. Panels (f)-(j) depict the outcomes of plumbene/H-half-saturated-BaSe(111) in the same manner, but with the light intensities chosen as $A$ =  0.005 Å$^{-1}$, 0.100 Å$^{-1}$, 0.165 Å$^{-1}$ and 0.190 Å$^{-1}$ respectively in LDOS patterns, in which the light frequency is selected as 0.475 eV. Besides, the ching balls stand for Se atoms.}
		\label{fig4:Substrate}
	\end{figure*}

	\subsection{Plumbene on Substrates}
	
	In addition to the analysis of free-standing plumbene under tensile biaxial stresses, the substrate plays a pivotal role in determining the ultimate topology. For simplicity, we focus exclusively on non-magnetic substrates. First, for substrates with vdW-layered characteristics, no bonding occurs with the dangling bonds of plumbene. Here we select h-BN as a template. h-BN serves as a highly suitable vdW substrate for plumbene \cite{hiramony2023numerical,gjerding2021recent}.  Considering the in-plane lattice constants of h-BN ($\textbf{\textit{a}}$ = 2.51 Å), the 2$\times$2 supercell of h-BN ($\textbf{\textit{a}}$ = 5.02 Å) fits well with the single cell of plumbene ($\textbf{\textit{a}}$ = 4.93 Å), resulting in only +1.8 \% tensile strain on the latter. This moderate tensile strain significantly aligns the band gap between the $K$ ($K'$) valleys and the $\Gamma$ point, making plumbene/h-BN an ideal building block for exhibiting the topological characteristics of tensile-strained free-standing plumbene, as discussed in Fig. 1. Figure S10 presents three primary stacking types of plumbene/h-BN, all of which maintain structural stability, while the band structure of plumbene itself remains largely unaffected.

	Figures \ref{fig4:Substrate}(a)-\ref{fig4:Substrate}(e) present the side view structure and light-engineered topological phase transitions of plumbene/h-BN in stacking type I. To facilitate direct comparison with free-standing plumbene under a tensile strain of +3.0 \%, the same laser parameters are applied. Notably, the three topological characteristics, Phases I, II, and III, exhibiting Chern numbers of $-8$, $-6$, and $-3$ respectively, remain unchanged. These properties have also been verified through AHC spectra and Wannier charge center distributions (Fig. S11). These intriguing results demonstrate that plumbene/h-BN provides an ideal and experimentally accessible method to explore the novel properties of tensile-strained free-standing plumbene under Floquet engineering.

	Conversely, the non-vdW substrate forms strong chemical bonds with plumbene, completely eliminating the $K$($K'$)-valley-based topology. To prevent the interference caused by the substrate's energy bands, we selected H-half-saturated BaSe(111) [see Fig. \ref{fig4:Substrate}(f)] \cite{jainmaterials}, which exerts a compressive stress of --4.3\% on plumbene. A portion of the Pb 6$p_\pm$ components in the valence band becomes more exposed compared to the lower distribution of Se 4$p$ components (see Fig. S12 in Supplementary Materials \cite{supplementary}), which is propitious to observing the phase transitions by persisting the global gap. Following this, Figs. \ref{fig4:Substrate}(g)-\ref{fig4:Substrate}(j) display the R-CPL-induced TPT of plumbene grown on H-half-saturated BaSe(111) [the structure illustrated with Fig. \ref{fig4:Substrate}(f)] at $\hbar\omega$ = 0.475 eV. Remarkably, the $K$($K'$)-valley-based QAH state is entirely suppressed, leaving only the $\Gamma$-based QAH state with a Chern number of +3 as the light intensity increases beyond 0.15 Å$^{-1}$ [also authenticated by the AHC spectrum and Wannier charge center distribution (Fig. S13)], with the Chern insulating gap further enlarging as the light intensity increases to higher values [Figs. \ref{fig4:Substrate}(i) and \ref{fig4:Substrate}(j)]. Before the emergence of $\Gamma$-based QAH state, the metallic phase remains around the $\Gamma$ point. Additionally, Fig. S14 in Supplementary Materials \cite{supplementary} provides a contour distribution of global gaps, uncovering that a small QAH region forms in the large light-intensity regime. The $K$($K'$)-valley-based QAH island is absolutely absent, replaced instead by a trivial insulating region in the low-frequency range.

\section{Discussion}

	The reversal of chirality of the incident light, specifically left-handed CPL, induces the same phase transition process but opposite signs of the Chern numbers in each phase. Oppositely, LPL does not support the QAH state; instead, it disrupts the band overlap, leading to a gapped structure. All these results are illustrated in Figs. S15 and S16 \cite{supplementary}.

	In the following subsections, we will separately discuss the theoretical models, highlight the novel aspects of our work, assess the experimental feasibility, and examine the heat effects on the sample.

\subsection{Theoretical Models}

	The multiple-mechanism based topology in plumbene offers abundant Floquet-QAH states that absent in most of previously reported systems. First of all, we give a model analysis towards the intriguing findings of in this work.

	The topological characteristics around the $K$ ($K'$) valleys and the $\Gamma$ point share completely different manifestations. Concretely, for the $K$ ($K'$) valleys, the Haldane model \cite{haldane1988model} accounts for its topological states, which can be further concretized by the Kane-Mele model \cite{kane2005quantum}:

	\begin{equation}
	H_K(\textbf{\textit{k}}) = v_{\mathrm F} \bigl( \tau_z k_x \sigma_x + k_y \sigma_y \bigr) + \lambda_{\text{SOC}} \sigma_z \tau_z s_z
	\end{equation}

	In eq. (6), $v_{\mathrm F}$, $\lambda_{\mathrm {SOC}}$, $\sigma_{x,y,z}$, $\tau_{x,y,z}$ and $s_{x,y,z}$ stand for the Fermi velocity, SOC effect and the Pauli matrices describing the sublattice space, valley space and spin space sequentially. For simplicity, we adopt the natural units ($\hbar$ = $c$ = $e$ = 1). In the absence of inversion symmetry breaking, the Rashba term can be neglected. The above low-energy effective model characterizes the --1 order of bottom conduction band and the +1 order of top valence band, which govern the TPT in this work.

	The Chern number of the above Kane-Mele model at each valley and spin render a half-integer value, taking the valley $K$ of spin up as example: \cite{kane2005quantum,bernevig2013topological,liu2016quantum}:

\begin{equation}
	\begin{split}
		C_K^\uparrow &= \frac{1}{4\pi} \iint \mathrm {d} k_x \, \mathrm {d} k_y \, 
		\frac{1}{\lvert \mathbf{d}(\textbf{\textit{k}}) \rvert^3}
		\mathbf{d}(\textbf{\textit{k}}) \cdot
		\left( \frac{\partial \mathbf{d}(\textbf{\textit{k}})}{\partial k_x} \times
		\frac{\partial \mathbf{d}(\textbf{\textit{k}})}{\partial k_y} \right) \\
		&= \frac{1}{4\pi} \iint k \, \mathrm {d}k \, \mathrm {d}\theta \,
		\frac{\lambda_{\text{SOC}}}{\left( v_{\mathrm F}^2 k^2 + \lambda_{\text{SOC}}^2 \right)^{3/2}}
		= \frac{1}{2} \operatorname{sgn}(\lambda_{\text{SOC}})
	\end{split}
\end{equation}

	Transforming the valley and spin indices:

\begin{equation}
	C_{K'}^\uparrow = C_K^\uparrow = -C_K^\downarrow = -C_{K'}^\downarrow
\end{equation}

	For each spin, although single valley contributes a half-integer Chern number, the integration of one band in whole BZ must provide a integer Chern number. Here, $C_{K^\prime}^\uparrow = C_K^\uparrow$, totally leading to $\pm$1.

	Conversely, for $\Gamma$ point, the effective model needs to be altered into the Qi-Wu-Zhang model \cite{qi2006topological,bernevig2006quantum,liu2016quantum}, shown in the continuum limit near the $\Gamma$ point:

	\begin{equation}
	\begin{split}
		H_{\mathrm{QWZ}}(\textbf{\textit{k}}) &= \eta J k_x \sigma_x - J k_y \sigma_y + \\ &\left[ \lambda_{\mathrm{SOC}} + M + B (k_x^2 + k_y^2) \right] \sigma_z
	\end{split}
	\end{equation}

	Here, $\eta = \pm 1$ stands for the spin up and down, respectively. The Chern number of Qi-Wu-Zhang model rigorously offers an integer, with "+" and "$-$" representing the conditions of spin up and down respectively:

	\begin{equation}
	C = \pm \frac{1}{2} \left[ \operatorname{sgn}(B) - \operatorname{sgn}(\lambda_{\mathrm{SOC}} + M) \right]
	\end{equation}

	Thus, the basis of the above two models is thoroughly different, meaning the two unalike mechanisms coexisting in plumbene.

	However, the most of band inversion points are not exactly located at the $K$, $K'$ and $\Gamma$ points, but distributed around these high symmetry points with $C_{3z}$ or $C_{6z}$ rotational symmetries. Moreover, the high Chern number emerges both around the $K$ ($K'$) valleys and the $\Gamma$ point. These phenomena indicate the appearance of the trigonal warping effect \cite{kechedzhi2007influence,rakyta2010trigonal,zeng2017topological,joucken2020determination,wu2021nonlinear,zhan2023floquet,akay2018trigonal,fu2009hexagonal,liu2013plane,ren2016quantum,li2022chern,xue2024tunable,wu2023robust,zhong2017plane}.

	For the $K$ ($K'$) valleys in graphene-like systems, the trigonal warping term is described by $H_{\mathrm W}=\lambda_{\mathrm W}\left[\left(k_x^2-k_y^2\right)\sigma_x-2k_xk_y\sigma_y\right]$ \cite{akay2018trigonal}. For the $\Gamma$ point, however, it is replaced by: $H_{\mathrm {W-FL}}=\frac{\lambda_{\mathrm W}}{2}\left(k_+^3+k_-^3\right)\sigma_zs_z=\lambda_{\mathrm W}\left(k_x^3-3{k_xk}_y^2\right)\sigma_zs_z$, which was firstly purposed by Prof. Liang Fu in 2009 \cite{fu2009hexagonal} and further investigated by X. Liu \textit{et al}. in 2013 \cite{liu2013plane}, Y. Ren \textit{et al}. in 2016 \cite{ren2016quantum}, Z. Li \textit{et al}. in 2022 \cite{li2022chern}, and so on \cite{xue2024tunable,wu2023robust,zhong2017plane}. The emergence of both the two forms of trigonal warping term yields the possibility of the band gap minimum not exactly at the Dirac point, but trigonally or hexagonally distributed around it. A trigonal or hexagonal band shape also forms in low energy regime.

	Directly determining the Chern number at each sub-valley corresponding to the band inversion point is challenging. Nonetheless, it is important to note that each sub-valley has a counterpart located at the spatial inversion point. According to previous study \cite{li2022chern}, these sub-valleys are analogous to the valleys described in the Kane-Mele model \cite{kane2005quantum}. Therefore, each sub-valley contributes a half-integer Chern number, resulting in a cumulative contribution of $\pm$ 3 for each spin component around the $K$ ($K'$) valleys or the $\Gamma$ point. 

	In plumbene, the --1 order conduction band and the +1 order valence band dominate the entire TPT process, while the zero-order bands play a negligible role. Consequently, the Floquet-Bloch Hamiltonian for this system requires a similarity transformation to a new spatial representation, where the --1 order conduction band and the +1 order valence band serve as the core bands, with the other bands acting as perturbations. The similar transformation has been purposed previously in semiconductor quantum wells \cite{lindner2011floquet}. Given that $\hbar\omega\ = 0.375$ eV is the primary light frequency utilized, we observe a slow convergence of the Magnus expansion \cite{liu2023floquet,blanes2009magnus,bukov2015universal,zhu2023floquet}. Although this approach may not be suitable for quantitative estimations, it provides a useful framework for qualitative analyses of the topological transformations predicted in plumbene. Detailed derivations can be found in Section 1 of the Supplementary Materials \cite{supplementary}.

	For Kane-Mele model plus trigonal warping term, under the irradiation of R-CPL: $\textbf{\textit{A}}=A(\cos\omega t,\ \ \sin\omega t,\ 0)$, the effective mass term gives the form:

	\begin{equation}
	M_{K,K'}(\textbf{\textit{k}}) = \left[ \frac{v_{\mathrm F}^2 A^2 s_z}{\omega} - \frac{4 A^2 \lambda_{\mathrm W}^2 k^2 s_z}{\omega} + M_{-1,1}(k) \right] \tau_z \sigma_z
	\end{equation}

	In eq. (11), $M_{-1,1}(k)$ is the “pseudo-SOC” term. At the point of $k = 0$, $M_{-1,1}=\left|\lambda_{\mathrm {SOC}}-2\omega\right|$. Additionally, we set $k^2=k_x^2+k_y^2$.

	For the Qi-Wu-Zhang model plus trigonal warping term, the effective mass term becomes:

	\begin{equation}
	\begin{split}
		M_{\mathrm{QWZ}}(\textbf{\textit{k}}) = \biggl[ & \lambda_{\mathrm{SOC}} - 2\omega + M + B(A^2 + k^2) - \frac{A^2 J^2 s_z}{\omega} \\ 
		& + \lambda_{\mathrm W} (k_x^3 - 3k_x k_y^2) s_z \biggr] \sigma_z
	\end{split}
	\end{equation}

	Markedly, the trigonal warping effect introduces an additional light-related mass term in the Kane-Mele model; however, in the Qi-Wu-Zhang model, it merely reshapes the two bands without adding an extra mass term.

	Finally, the eigenvalues around the $K$ ($K'$) valleys yield:

	\begin{equation}
	\begin{split}
		E_{K,K' \pm} &= \pm \biggl\{ v_{\mathrm F}^2 k^2 \pm 2\eta v_{\mathrm F}^2 \lambda_{\mathrm W} k^3 \cos 3\theta  + v_{\mathrm F}^2 \lambda_{\mathrm W}^2 k^4 \\
		& + \left( \frac{\eta v_{\mathrm F}^2 A^2}{\omega} - \frac{4\eta A^2 \lambda_{\mathrm W}^2 k^2}{\omega} + M_{-1,1}(k) \right)^2 \biggr\}^{1/2}
	\end{split}
	\end{equation}

	In eq. (13), $k_x = k\cos\theta$, $k_y = k\sin\theta$, $\eta = \pm 1$ corresponds to the spin up and down. Notably, the effective mass terms introduced by the Dirac-band term ($\frac{{v_{\mathrm F}^2A}^2}{\omega}$) and the trigonal warping term ($\frac{4A^2\lambda_W^2k^2}{\omega}$) have opposite signs, indicating that they contribute to opposing chirality in the Chern number jumps generated by these two terms. This characteristic aligns with the observed behavior of Chern number transitions at the Dirac point, where there is a jump of +2, as well as those occurring near the Dirac point, which transition from a metallic state to a negative value.

	Considering the Diarc point ($k = 0$), eq. (13) is reduced into:

	\begin{equation}
	{E_{K,K' \pm}} = \pm \left( \frac{\eta v_{\mathrm F}^2 A^2}{\omega} + M_{-1,1}(k) \right)
	\end{equation}

	Spin up does not undertake a band closure, with the band gap increasing monotonically [same to red band curves depicted in Figs. \ref{fig2:Detail_analysis}(a)-\ref{fig2:Detail_analysis}(e)]. In the contrary, spin down closes the gap at $\frac{{v_{\mathrm F}^2A}^2}{\omega}=M_{-1,1}\left(k\right)$, with a positive Chern-number jump at both valleys [same to blue band curves depicted in Figs. \ref{fig2:Detail_analysis}(a)-\ref{fig2:Detail_analysis}(e)]. Therefore, this result is completely consistent with those computational results shown in Figs. \ref{fig2:Detail_analysis}(a)-\ref{fig2:Detail_analysis}(e).

	Then, when we fix the value of $k$, eq. (13) is reduced to:

	\begin{equation}
	{E_{K,K'\pm} = \pm \sqrt{P + Q \pm \eta R \cos 3\theta}}
	\end{equation}

	$\theta$ is the only variable in eq. (15). Finding the band gap minimum gives the relationship: $\cos3\theta = \pm 1$, which further offers:

	\begin{equation}
	\theta = 
	\begin{cases}
		\dfrac{(2n+1)\pi}{3}, & \text{for } K(\uparrow) \text{ and } K'(\downarrow),\ n \in \mathbb{Z} \\
		\dfrac{2n\pi}{3}, & \text{for } K(\downarrow) \text{ and } K'(\uparrow),\ n \in \mathbb{Z}
	\end{cases}
	\end{equation}

	Eq. (16) successfully explains the $C_{3z}$ rotational symmetry based band gap minimums around the $K$ ($K'$) valleys. In like manner, the Qi-Wu-Zhang model forms the light-induced effective model, offering the eigenvalues in the following:

	\begin{equation}
	\begin{split}
		E_\pm = \pm \biggl\{ & \alpha^2 k^2 + 6\eta\alpha\beta (k_x^3 - 3k_x k_y^2) + 9\beta^2 k^4 \\
		& + \left[ M_{\mathrm{eff}}^{\uparrow\downarrow} + \lambda_{\mathrm W} \eta (k_x^3 - 3k_x k_y^2) \right]^2 \biggr\}^{1/2}
	\end{split}
	\end{equation}

	In eq. (17), we define $\alpha=J+\frac{2BA^2}{\omega}$, $\beta=\frac{\lambda_{\mathrm W}JA^2}{\omega}$, $M_{\mathrm {eff}}^{\uparrow\downarrow}=\lambda_{\mathrm {SOC}}-2\omega+M+B{(A^2+k}^2)\mp\frac{A^2J^2}{\omega}$. For the spin-up band, band closure occurs when $\frac{A^2J^2}{\omega}$ reaches the threshold given by $\lambda_{\mathrm {SOC}}-2\omega+M+BA^2$. This closure begins at the $\Gamma$ point and subsequently expands to finite momentum, resulting in a sixfold distribution property. This near-zero band gap persists for light intensities ranging from 0.125 Å$^{-1}$ $\sim$ 0.135 Å$^{-1}$ [see Fig. \ref{fig3:Contours}(a), where the photon energy is 0.375 eV], after which the band gap increases significantly. Due to the constraints of the model, eq. (10) only exhibits $C_{3z}$ rotational symmetry. Nonetheless, due to inversion symmetry, each minimum gap point has a counterpart at the same momentum radius but with the opposite value of $\theta$, resulting in a sixfold band inversion characteristic.

	All of the derivations and detailed discussions of the theoretical models, along with their physical implications, can be found in Section 1 of the Supplementary Materials \cite{supplementary}.

\subsection{Novelty in Floquet-engineered Plumbene}

	As a summary, there exist three aspects of novelty in Floquet-engineered topology, which position plumbene as a promising system for future investigations.

	Firstly, in most prior theoretical proposals, Floquet-governed TPTs are typically described by a single mechanism. For example, in the cases of silicene, germanene, and stanene \cite{li2025light}, the TPT occurs once, transitioning from a QSH state to a QAH state driven by the Kane-Mele model \cite{kane2005quantum}, without contributions from other factors. Additionally, monolayer ScI$_2$ and VSNH undergo a two-stage TPT process \cite{li2024floquet,tian2025quantized}, first transitioning from a high-order topological insulator to a QAH state, and then to a trivial state, though still describable by a similar valley-based model in an inversion-symmetry-broken regime. Monolayer SMoSiN$_2$ was also predicted to perform a high Chern number QAH state under the irradiation of CPL \cite{feng2025engineering}, but its high Chern number is solely contributed from the $K$($K'$) valley with polarity, and is fixed at $\pm$ 6 without tunability. Furthermore, the Floquet-engineered quasi-topological surface states of Bi$_2$Te$_3$ or MnBi$_2$Te$_4$ films \cite{zhu2023floquet,qin2023light} also adhere to single Dirac-cone-like models. In contrast, plumbene exhibits nontrivial topology at both the $K$ ($K'$) valleys and the $\Gamma$ point, characterized by the Kane-Mele model \cite{kane2005quantum} and the Qi-Wu-Zhang model \cite{qi2006topological,bernevig2006quantum,liu2016quantum}, respectively. Moreover, the significant difference between the energy gaps at the $K$ ($K'$) valleys and the $\Gamma$ point leads to the non-simultaneous occurrence of topological phase transitions, resulting in a richer emergence of various high Chern numbers.

	Secondly, In Dirac cone-like systems, the trigonal warping effect plays a fundamental role in reshaping the low-energy continuum band models near the Dirac point. To date, only monolayer VSi$_2$N$_4$ has been predicted to develop a trigonal warping effect in one valley under specific laser parameters \cite{zhan2023floquet}. Plumbene exhibits a large band gap in its ground state, allowing the $\pm$ 1 replica bands to dominate the entire TPT process in the low-frequency regime considered in this work. For the $K$ ($K'$) valleys, the trigonal warping effect is present throughout the entire process of topological evolution due to initial band overlapping, resulting in a Chern number of $-$6. At the $\Gamma$ point, the trigonal warping effect maintains a Chern number of +3 following band inversion. Moreover, under the irradiation of CPL, the trigonal warping effect contributes effective mass terms in the Kane-Mele model \cite{kane2005quantum} but only reshapes the Dirac bands within the Qi-Wu-Zhang model \cite{qi2006topological,bernevig2006quantum,liu2016quantum}. These novel findings inspire future research on Floquet-engineered trigonal warping effects, revealing previously unexplored phenomena.

	Thirdly, the trigonal warping term can coexist with Chern numbers linked to Dirac cones. In Phase I of free-standing plumbene and plumbene/h-BN, the Chern number of --4 in each valley arises from the trigonal warping effect (--3) and the Dirac model (--1). While previous models predicted this coexistence \cite{zeng2017topological}, it is the first observation in real materials driven by CPL. Additionally, the light-induced effective mass terms from both models have opposite signs, suggesting they evolve in opposing directions, which may enhance the emergence of further Chern numbers in this system.

\subsection{Experimental Feasibility}

	The light frequencies used in this work range from $\hbar\omega$ = 0.2 eV to 0.6 eV, corresponding to the middle infrared wavelength region. Several laser options can achieve this range with continuous frequency tunability, including titanium-sapphire lasers and optical parametric oscillators.

	While titanium-sapphire lasers allow tunability across red visible to near-infrared wavelengths, obtaining middle infrared light requires the difference-frequency generation method, which mixes two light frequencies in a nonlinear optical crystal. Optical parametric oscillators are also effective for generating widely tunable, coherent middle infrared radiation \cite{tang1992optical,harris1969tunable}. This method uses nonlinear optical crystals and can be adjusted by tuning the pump beam's propagation axis or changing the refractive index through temperature. Thus, middle infrared light with photon energies from 0.2 eV to 0.6 eV can be smoothly obtained.

	To generate CPL, linearly polarized light (LPL) can be produced using a standard polaroid filter. When LPL is incident at a 45$^{\circ}$ angle to the optical axis of a quarter-wave plate, it generates CPL, and rotating the angle by 90$^{\circ}$ reverses the chirality of the CPL.

\subsection{Estimation of the Heat Effect}

	The thermal effect is another critical factor in discussions of Floquet engineering. Unfortunately, there is no existing research that directly verifies the precise threshold for structural breakdown in plumbene, or even in lead itself. Generally, the fluence of the light is derived from the Poynting vector of the electromagnetic field, which is expressed as follows:

	\begin{equation}
	F = I t = \frac{1}{2} \varepsilon_0 c \omega^2 A_0^2 \times \frac{2\pi N}{\omega} = \pi \varepsilon_0 c \omega N A_0^2
	\end{equation}

	In eq. (18),  $\varepsilon_0$, $c$, $\omega$, $N$ and $A_0$ stand for the vacuum permittivity, light velocity, angular frequency of the light, the numbers of cycles, and light intensity respectively, with all of them in international units. Based on comparable studies on group-IV Xenes \cite{li2025light}, the maximum fluence reported is 0.013 J/cm$^{2}$. Six cycles of laser irradiation are sufficient for first-order replica-band generation \cite{liu2018photoinduced,qiu2018ultrafast}. As indicated in this work, the maximum laser frequency and amplitude used are $\hbar\omega$ = 0.6 eV and $A$ = 0.20 Å$^{-1}$, respectively, resulting in a maximum fluence of 0.0079 J/cm$^{2}$.

	For direct comparison, we conducted first-principles computations to estimate the bond formation energies of plumbene in comparison to silicene, germanene, and stanene, with the results shown in Fig. S17 \cite{supplementary}. The bond formation energies are --9.38 eV/atom for silicene, --7.54 eV/atom for germanene, and --5.69 eV/atom for stanene, while plumbene exhibits a weaker bonding strength at --3.14 eV/atom. Referring to the damage thresholds established for Si and Ge in Refs. \cite{li2025light,sokolowski2001femtosecond}, and considering the ratio of bond formation energies, the damage threshold for Pb is estimated to be around 0.02 J/cm$^{2}$, which is higher than the maximum fluence used in this study. Consequently, the laser parameters employed in this work are expected to avoid thermal damage.

	\section{Conclusion}

	In this work, we utilize Floquet engineering to uncover the hidden topological characteristics of plumbene, which initially behaves as a large gapped high-order topological insulator. When irradiated by R-CPL and with laser-amplitude enhancement, a three-stage TPT process occurs, with each QAH phase characterized by high Chern numbers: --8, --6, and --3, respectively. Markedly, we firstly purposed that sufficiently strong light intensities induce a mQAH state in the pure material, plumbene, yielding a compensated Chern number as --3, distinct from germanene/MnBi$_2$Te$_4$, where different QAH state is individually contributed by different layer components \cite{li2024multimechanism}.
	
	To achieve the necessary band gap alignment for these three QAH states, moderate tensile strains are essential for obtaining global gaps, which are critical for transport measurements. The contour distributions of global gaps and Chern numbers indicate a two-QAH-island feature, with a large gapped region ($\geq$ 40 meV) located within each island. A vdW substrate preserves most of the topological characteristics of plumbene while compromising its valley-polarized nature, whereas a non-vdW substrate forms bonds with plumbene, filtering out the $K$($K'$)-valley-based QAH state and resulting in a $\Gamma$-point-based QAH state with $C$ = +3. The insights gained from this work enhance our understanding and design of next-generation dissipationless electronic devices.

	

	\begin{acknowledgments}
		We thank Prof. Feng Liu, Dr. Hui Zhou, Dr. Xiyu Hong and Wenlu Zhang for helpful discussions, and thank Asso. Prof. Huixia Fu, Dr. Hang Liu for technical supports. The numerical calculations have been done on the supercomputing system in the Huairou Materials Genome Platform. This work is supported by National Natural Science Foundation of China (Grants Nos. 12025407, 12450401 and 12274276), National Key Research and Development Program of China (Grant No. 2021YFA1400201), Chinese Academy of Sciences (Grants Nos. YSBR-047 and XDB33030100), and the Natural Science
		Foundation of Shanxi Province (China) (Grant No. 202403021223008).  
	\end{acknowledgments}

	\newpage
	\nocite{*}

\end{document}